\documentclass[twoside,twocolumn,9pt]{article}
\usepackage{extsizes}
\usepackage[super,sort&compress,comma]{natbib} 
\usepackage[version=3]{mhchem}
\usepackage[left=1.5cm, right=1.5cm, top=1.785cm, bottom=2.0cm]{geometry}
\usepackage{balance}
\usepackage{mathptmx}
\usepackage{sectsty}
\usepackage{graphicx} 
\usepackage{lastpage}
\usepackage[format=plain,justification=justified,singlelinecheck=false,font={stretch=1.125,small,sf},labelfont=bf,labelsep=space]{caption}
\usepackage{float}
\usepackage{fancyhdr}
\usepackage{fnpos}
\usepackage[english]{babel}
\usepackage{amssymb}
\usepackage{amsmath}
\usepackage{bm}
\DeclareMathOperator{\Tr}{Tr}
\addto{\captionsenglish}{%
  \renewcommand{\refname}{Notes and references}
}
\usepackage{array}
\usepackage{droidsans}
\usepackage{charter}
\usepackage[T1]{fontenc}
\usepackage[usenames,dvipsnames]{xcolor}
\usepackage{setspace}
\usepackage[compact]{titlesec}
\usepackage{hyperref}

\usepackage{epstopdf}

\definecolor{cream}{RGB}{222,217,201}

\begin{document}

\pagestyle{fancy}
\thispagestyle{plain}
\fancypagestyle{plain}{
\renewcommand{\headrulewidth}{0pt}
}

\makeFNbottom
\makeatletter
\renewcommand\LARGE{\@setfontsize\LARGE{15pt}{17}}
\renewcommand\Large{\@setfontsize\Large{12pt}{14}}
\renewcommand\large{\@setfontsize\large{10pt}{12}}
\renewcommand\footnotesize{\@setfontsize\footnotesize{7pt}{10}}
\makeatother

\renewcommand{\thefootnote}{\fnsymbol{footnote}}
\renewcommand\footnoterule{\vspace*{1pt}%
\color{cream}\hrule width 3.5in height 0.4pt \color{black}\vspace*{5pt}} 
\setcounter{secnumdepth}{5}

\makeatletter 
\renewcommand\@biblabel[1]{#1}            
\renewcommand\@makefntext[1]%
{\noindent\makebox[0pt][r]{\@thefnmark\,}#1}
\makeatother 
\renewcommand{\figurename}{\small{Fig.}~}
\sectionfont{\sffamily\Large}
\subsectionfont{\normalsize}
\subsubsectionfont{\bf}
\setstretch{1.125} 
\setlength{\skip\footins}{0.8cm}
\setlength{\footnotesep}{0.25cm}
\setlength{\jot}{10pt}
\titlespacing*{\section}{0pt}{4pt}{4pt}
\titlespacing*{\subsection}{0pt}{15pt}{1pt}

\fancyfoot{}
\fancyfoot[RO]{\footnotesize{\sffamily{1--\pageref{LastPage} ~\textbar  \hspace{2pt}\thepage}}}
\fancyfoot[LE]{\footnotesize{\sffamily{\thepage~\textbar\hspace{3.45cm} 1--\pageref{LastPage}}}}
\fancyhead{}
\renewcommand{\headrulewidth}{0pt} 
\renewcommand{\footrulewidth}{0pt}
\setlength{\arrayrulewidth}{1pt}
\setlength{\columnsep}{6.5mm}
\setlength\bibsep{1pt}

\makeatletter 
\newlength{\figrulesep} 
\setlength{\figrulesep}{0.5\textfloatsep} 

\newcommand{\topfigrule}{\vspace*{-1pt}%
\noindent{\color{cream}\rule[-\figrulesep]{\columnwidth}{1.5pt}} }

\newcommand{\botfigrule}{\vspace*{-2pt}%
\noindent{\color{cream}\rule[\figrulesep]{\columnwidth}{1.5pt}} }

\newcommand{\dblfigrule}{\vspace*{-1pt}%
\noindent{\color{cream}\rule[-\figrulesep]{\textwidth}{1.5pt}} }

\makeatother

\twocolumn[
  \begin{@twocolumnfalse}
\vspace{1em}
\sffamily
\begin{tabular}{m{4.5cm} p{13.5cm} }

& 
 \noindent\LARGE{\textbf{Equilibrium morphology of tactoids in elastically anisotropic nematics}} \\
\vspace{0.3cm} & \vspace{0.3cm} \\

& \noindent\large{Cody D. Schimming* and Jorge Vi\~nals} \\

& \noindent\normalsize{We study two dimensional tactoids in nematic liquid crystals by using a $\mathbf{Q}$-tensor representation. A bulk free energy of the Maier-Saupe form with eigenvalue constraints on $\mathbf{Q}$, plus elastic terms up to cubic order in $\mathbf{Q}$ are used to understand the effects of anisotropic anchoring and Frank-Oseen elasticity on the morphology of nematic-isotropic domains. Further, a volume constraint is introduced to stabilize tactoids of any size at coexistence. We find that anisotropic anchoring results in differences in interface thickness depending on the relative orientation of the director at the interface, and that interfaces become biaxial for tangential alignment when anisotropy is introduced. For negative tactoids, surface defects induced by boundary topology become sharper with increasing elastic anisotropy. On the other hand, by parametrically studying their energy landscape, we find that surface defects do not represent the minimum energy configuration in positive tactoids. Instead, the interplay between Frank-Oseen elasticity in the bulk, and anisotropic anchoring yields semi-bipolar director configurations with non-circular interface morphology. Finally, we find that for growing tactoids the evolution of the director configuration is highly sensitive to the anisotropic term included in the free energy, and that minimum energy configurations may not be representative of kinetically obtained tactoids at long times.} \\

\end{tabular}

 \end{@twocolumnfalse} \vspace{0.6cm}

  ]

\renewcommand*\rmdefault{bch}\normalfont\upshape
\rmfamily
\section*{}
\vspace{-1cm}


\footnotetext{\textit{School of Physics and Astronomy, University of Minnesota, Minneapolis, Minnesota 55414, USA. E-mail: schim111@umn.edu}}




\section{Introduction}
Liquid crystals are materials composed of anisotropic constituents. This, in turn, leads to anisotropic optical and hydrodynamic properties in the ordered nematic phase \cite{deGennes75}. Moreover, effects of the underlying anisotropy can even be observed at the first-order isotropic-nematic phase transition. Indeed, modelling of spindle-shaped nematic domains, or ``tactoids,'' continues to pose a theoretical challenge.\cite{prinsen03,prinsen04,prinsen04b,wincure06,vanderschoot12,zhang18,golovaty20,paparini21,paparini21b} Understanding the effects of surface tension, surface anchoring, and elastic anistropy on the morphology of tactoids is important to predicting the anisotropic structure of many engineered and biological materials, where the value of these parameters can vary markedly.\cite{lopez11,lazo14,peng15,patteson18,peng18,baba18,hokmabad19,coelho20,lee20,ludwig20}

There have been many recent analytical and computational studies of the equilibrium shape and structure of ``positive'' tactoids (domains of nematic phase in an isotropic matrix).\cite{prinsen03,prinsen04,prinsen04b,wincure06,vanderschoot12,everts16,zhang18,paparini21,paparini21b} Such domains nucleate spontaneously during the isotropic-nematic phase transition in the coexistence region of the phase diagram.\cite{kaznacheev02,kim13,jamali15} ``Negative'' tactoids (domains of isotropic phase in a nematic matrix) have also been observed experimentally,\cite{kim13,zhou17} typically forming from the melting of disclination cores, although there are fewer theoretical studies of these structures.\cite{golovaty20} Further, the effect of material parameters on domain growth is not common in the theoretical literature.\cite{wincure06}

The preferred method for modelling positive tactoids has been through the use of the nematic director, $\mathbf{n}$, a unit vector representing the local orientation of nematogens. The simplicity of this method allows for analytical results when the shape of tactoid is fixed, and the director configuration is frozen.\cite{prinsen03,prinsen04,prinsen04b} Further, computational results can clarify the equilibrium structure of the director if the tactoid shape remains fixed.\cite{vanderschoot12} Often the goal of such studies is to find the shape and director configuration that minimizes the bulk and surface energies. Methods that allow for the boundary to change, so as to allow the system to relax to a true equilibrium shape, are just starting to surface \cite{debenedictis16}. However, the isotropic matrix cannot be modelled by such a method. The interface is therefore treated as a surface on the edge of the domain, rather than a diffuse structure as seen in experiments.\cite{kim13} To model the interface, a scalar order parameter, $S$, must be introduced that characterizes the degree of ordering of the constituent molecules. Thus when $S=0$, the system is in an isotropic phase.

The isotropic-nematic interface can also be resolved by use of a tensor order parameter $\mathbf{Q}$. The eigenvalues of $\mathbf{Q}$ represent the degree of ordering while the eigenvectors represent the nematogen orientation. Here the zero tensor represents the isotropic phase. In addition to being able to resolve a diffuse interface between phases, the $\mathbf{Q}$ tensor allows for biaxial order, which has been shown to appear in computational studies of interfaces using a Landau-de Gennes free energy, the typical free energy associated with $\mathbf{Q}$.\cite{popanita97} Additionally, the $\mathbf{Q}$-tensor approach may be used to resolve defects on the surface or interior of tactoids. The challenges with working with $\mathbf{Q}$ are that one needs to solve tensor valued equations, leading to more computationally complex systems. In addition, when comparing elastic modes of the Landau-de Gennes free energy to those of Frank-Oseen (the typical elastic energy associated with the director\cite{frank58,selinger16}) one must go to at least third order in $\mathbf{Q}$ to individually resolve the elastic modes. In most cases, this leads to an unbounded free energy, and, hence, to an ill-posed problem.\cite{longa87,ball10,schimming21}

In this work, we extend the theoretical and computational literature on interfaces and tactoids by using a Maier-Saupe energy and a singular bulk free energy, function of $\mathbf{Q}$, originally presented by Ball and Majumdar, and recently developed computationally and conceptually.\cite{ball10,schimming20,schimming21} Using this free energy, one can adjust the Frank-Oseen elastic constants, and study their effect on interface and tactoid morphology and director structure. We also develop a computational technique to constrain the volume of quasi-2D tactoids which allows us to stabilize and study negative and positive tactoids of fixed volume. The paper is organized as follows:  In section 2, we summarize the model, including the fixed volume condition for tactoids, and analyze the expected effects of the proposed elastic energy. In section 3 we numerically study the one-dimensional isotropic-nematic interface and analyze the width and interfacial energies as a function of anchoring. In section 4 we numerically compute negative tactoids of character $\pm 1/2$ and discuss the emergence of cusps as the elasticity becomes more anisotropic. In section 5 we compute positive tactoids and study the energy landscape as a function of bipolarity of the director field and aspect ratio of the tactoid shape. Finally, in section 6 we look at tactoid growth from small positive tactoids to full nematic domains.

\section{Model}
\subsection{Maier-Saupe Self Consistent Field Theory}
We first briefly review the singular potential method of Ball and Majumdar\cite{ball10}, and its more recent computational implementation in Refs.\cite{schimming20,schimming21}. To computationally study nematic domains, we seek to minimize a free energy over a domain $\Omega$ given by
\begin{equation} \label{eqn:freeE}
    F = \int_{\Omega} \left[ f_B(\mathbf{Q}) + f_e(\mathbf{Q},\mathbf{\nabla}\mathbf{Q}) \right]\, d\mathbf{r}
\end{equation}
where $f_B$ is a bulk free energy density and $f_e$ is an elastic free energy density that penalizes spatial variations of the order parameter. Typically, $f_B$ is chosen to be the standard double-well Landau-de Gennes free energy.\cite{deGennes75,mottram14} 

The method of Ball and Majumdar builds upon the Maier-Saupe model \cite{maier59} instead. Consider an ensemble of nematogens, each described by a unit vector $\mathbf{u}$ giving its molecular orientation. Let $p(\mathbf{u})$ be the equilibrium probability density of orientations at constant temperature. The energy per unit volume is then given by
\begin{equation}
    E = -\kappa \int_{{\cal S}^{2}} \int_{{\cal S}^{2}} \left[ (\mathbf{u}\cdot\mathbf{u}')^{2}  - \frac{1}{3} \right] p(\mathbf{u}) p(\mathbf{u}') d\Sigma (\mathbf{u}) d\Sigma (\mathbf{u}') ,
\end{equation}
where integration is conducted over the unit sphere, and $\kappa$ is a constant interaction parameter. The entropy functional (per unit volume) relative to the isotropic phase is given by
\begin{equation} \label{eqn:ent}
    \Delta S = - n \int_{{\cal S}^2} p(\mathbf{u}) \ln\left[4 \pi p(\mathbf{u})\right] \, d \Sigma(\mathbf{u}),
\end{equation}
where Boltzmann's constant has been set to unity, and $n$ is the number density of nematogens. The bulk free energy is therefore given by 
\begin{equation} \label{eqn:BulkE}
    f_B = E - T \Delta S,
\end{equation}
where $T$ is the temperature (in units such that $k_B = 1$). Macroscopic orientational order is described by the tensor
\begin{equation}
\label{eqn:Qdef}
    \mathbf{Q} = \int_{{\cal S}^2} \left(\mathbf{u} \otimes \mathbf{u} - \frac{1}{3} \mathbf{I}\right) p(\mathbf{u}) \, d\Sigma( \mathbf{u}).
\end{equation}
Note that $\mathbf{Q}$ is defined as a thermal average. 

The singular potential method introduced by Ball and Majumdar\cite{ball10} now proceeds as follows. Specify a value of $\mathbf{Q}$ (or locally, if a field $\mathbf{Q}(\mathbf{r})$ is specified), and find the probability distribution function $p^{*}$ that minimizes $f_{B}$ over a set of admissible distributions for the given $\mathbf{Q}$. The result is
\begin{equation}
\label{eqn:Zdef}
    p^{*}(\mathbf{u}) = \frac{1}{Z} e^{[\mathbf{u}^{T} \bm{\Lambda} \mathbf{u}]} \quad {\rm with} \quad Z = \int_{{\cal S}^2} e^{[\mathbf{u}^T \bm{\Lambda} \mathbf{u}]} \, d\Sigma(\mathbf{u}).
\end{equation}
where $\bm{\Lambda}$ is a tensor-valued Lagrange multiplier associated with the constraint, Eq. \eqref{eqn:Qdef}.  The value of $\bm{\Lambda}$ can be determined by the self-consistency equation
\begin{equation} \label{eqn:SelfConsist}
    \frac{\partial \ln Z}{\partial \bm{\Lambda}} = \mathbf{Q} + \frac{1}{3} \mathbf{I}.
\end{equation}
$\bm{\Lambda}$ is then treated as a function of $\mathbf{Q}$, implicitly determined by Eq. \eqref{eqn:SelfConsist}. With this minimizer $p^{*}$, the bulk free energy density can then be completely expressed as a function of $\mathbf{Q}$ only: $f_B(\mathbf{Q}) = n T [\bm{\Lambda} : (\mathbf{Q} + 1 / 3 \mathbf{I}) + \ln 4 \pi - \ln Z] - \kappa \Tr(\mathbf{Q}^2)$.

The self consistent theory derived, and in particular the constraint (\ref{eqn:Qdef}), restricts the eigenvalues of $\mathbf{Q}$ to remain in the physically admissible range $-1/3 \leq \lambda_i \leq 2/3$. If the eigenvalues approach the range boundary, $f_B$ goes to infinity. This has important consequences for the possible choices of $f_e$ in Eq. \eqref{eqn:freeE}. 

In our current study, the following functional form of the elastic free energy is used
\begin{multline} \label{eqn:ElasticE}
    f_e(\mathbf{Q},\mathbf{\nabla}\mathbf{Q}) = L_1 \partial_k Q_{ij} \partial_k Q_{ij} \\
    + L_3 Q_{k\ell} \partial_k Q_{ij} \partial_{\ell} Q_{ij} + L_4 Q_{k\ell}\partial_i Q_{k\ell} \partial_j Q_{ij}
\end{multline}
where $\partial_i$ denotes $ \partial/\partial x_i$, and summation over repeated indices is assumed. The cubic term proportional to $L_{3}$ is the lowest order term in powers of $\mathbf{Q}$ that allows for anisotropic elasticity. When this functional form is compared with the Frank-Oseen elastic energy density, one finds that $K_{33} - K_{11} = 2 L_{3} S_{N}^{3}$ \cite{longa87,zhou17}, where $K_{11}$ and $K_{33}$ are the splay and bend elastic modulii respectively, and $S_{N}$ is the bulk equilibrium value of the nematic degree of order (defined below).

If the bulk free energy $f_{B}$ were the Landau-de Gennes free energy, when combined with the elastic energy \eqref{eqn:ElasticE} up to third order in $\mathbf{Q}$, the resulting functional $F$ is not bounded below. In this case, one must either remain at quadratic order in $f_{e}$, or go to quartic order for the full free energy to be bounded below.\cite{longa87} Since there are 14 different quartic order terms allowed by symmetry, the theory becomes intractable. On the other hand, the eigenvalue constraint in the singular bulk potential $f_{B}$ yields a finite functional $F$ with $f_{e}$ given in Eq. \eqref{eqn:ElasticE}. Thus the singular potential method allows consideration of anisotropic elasticity without having to resort to poorly understood quartic terms in the expansion for $f_{e}$

For interfaces and tactoids, there are three primary energies involved: excess surface energy (surface tension), energy associated with surface anchoring, and energy deriving from bulk elasticity. The surface tension gives the energy of a surface per unit area, the surface anchoring gives the energy of relative orientation of nematogens at the interface, and bulk elasticity refers to the elastic modes of the director, given by the Frank-Oseen elastic energy. These various contributions can be associated with gradients of $\mathbf{Q}$ if $\mathbf{Q}$ is parameterized as $\mathbf{Q} = S (\mathbf{n} \otimes \mathbf{n} - (1/3)\mathbf{I})$, where $S$ represents the degree of order, and $\mathbf{n}$ is the director. Energies associated with gradients of $\mathbf{Q}$ can then be categorized by their contribution to $|\mathbf{\nabla} S|^2$, or surface tension; $(\mathbf{n} \cdot \mathbf{\nabla}S)^2$, or surface anchoring; and $(\mathbf{\nabla} \cdot \mathbf{n})^2$, $(\mathbf{n} \cdot \mathbf{\nabla} \times \mathbf{n})^2$, $|\mathbf{n} \times (\mathbf{\nabla} \times \mathbf{n})|^2$ which represent splay, twist, and bend respectively. Since we will focus on one and two dimensional configurations, the twist term will always be zero. With our choice of $f_{e}$, the contribution to surface tension is $(2 L_1 - (2/9) S L_3 - (2/9) S L_4)|\mathbf{\nabla}S|^2$; the contribution to surface anchoring is $((2/3) S L_3 + (2/3) S L_4) (\mathbf{n}\cdot\mathbf{\nabla}S)^2$; and the contribution to bulk elastic energy is $(2 L_1 S^2 - (2/3) S^3 L_3)(\mathbf{\nabla}\cdot\mathbf{n})^2 + (2 L_1 S^2 + (4/3) S^3 L_3)|\mathbf{n}\times(\mathbf{\nabla}\times\mathbf{n})|^2$. Note that the $L_4$ term does not contribute to bulk elasticity, so it is a useful term in controlling the magnitude of surface anchoring without changing any bulk properties of the nematic.

To compute equilibrium configurations we minimize the free energy, Eq. \eqref{eqn:freeE}, by using a gradient flow method with a semi-implicit convex splitting algorithm \cite{schimming21}. The finite element MATLAB package FELICITY \cite{walker18} is used to compute the finite element matrices for the configurations. The gradient flow equations (Eq. \eqref{eqn:gradFlow}) are discretized on a triangular mesh and then solved for each time step, with fixed $\Delta t$. Further details of each calculation will be given later for the various configurations computed. For the rest of the paper we work with dimensionless quantities given by
\begin{gather*}
    \Tilde{f} = \frac{f}{n T}, \quad \Tilde{\mathbf{r}} = \frac{\mathbf{r}}{\xi}, \quad \xi^2 = \frac{L_1}{n T}, \quad \Tilde{L}_{i\neq 1} = \frac{L_i}{L_1} \\ 
    \Tilde{t} = \frac{t}{\tau}, \quad \tau = \frac{\gamma}{n T}
\end{gather*}
where the length scale $\xi$ is defined by the value of $L_1$, and will be set to unity in all computations. Dimensionless time is used for the gradient flow equations where the time scale $\tau$ is set by the value of $\gamma$, the rotational viscosity, which will also be set to unity in computations. Additionally, we will use the dimensionless parameter $\kappa / n T$ to control the phase behavior of the configuration. The tilde on all quantities is henceforth dropped for brevity.

\subsection{Volume Constraint}
An important variable in the determination of the equilibrium morphology of tactoids is their volume. This is a simple matter in studies in which the boundary is fixed. However, since the interface in a $\mathbf{Q}$-tensor description occurs naturally, and the order parameter is non-conserved, it is not possible to constrain the size of tactoids directly. One possible avenue is to couple a conserved phase field to the nematic and isotropic phases, that varies from one phase to the other.\cite{brand87,lowen10} One obvious candidate would be a conserved density field. However, in systems of chromonic liquid crystals, the primary motivation of this study, the density difference between phases is very small.\cite{kim13} Additionally, mass transport has not been reported in these experiments. Thus, it appears to us that this solution is computationally more complex with no added physical benefit. Here, instead, we implement a method for fixing the volume of a two-dimensional tactoid by directly using a volume constraint with its associated Lagrange multiplier.

First, note that the matrix
\begin{equation} \label{eqn:Adef}
    \mathbf{A} = \frac{3}{S_N}\begin{pmatrix}
                    1 & 0 & 0 \\
                    0 & 1 & 0 \\
                    0 & 0 & 0
                 \end{pmatrix}
\end{equation}
where $S_N$ is the value of $S$ in the nematic phase, has the property that $\Tr(\mathbf{A} \mathbf{Q}) = 1$ if the system is in the uniaxial nematic phase, and $\Tr(\mathbf{A} \mathbf{Q}) = 0$ if the system is in the isotropic phase. Further, $\mathbf{A}$ has the same form, given in Eq. \eqref{eqn:Adef}, whether we work in the eigenframe of $\mathbf{Q}$, or the fixed frame where $\mathbf{n}$ lies in the $xy$ plane. Thus we have
\begin{equation} \label{eqn:VolConstraint}
    \int_{\Omega}\Tr(\mathbf{A} \mathbf{Q}) \, d\mathbf{r} = V_N
\end{equation}
where $V_N$ denotes the volume of the nematic phase in the system. Equation \eqref{eqn:VolConstraint} serves as a constraint on the tactoid volume, and is implemented with a Lagrange multiplier $\mu$. The new free energy is $F^* = F - \mu \left[\int_\Omega \Tr(\mathbf{A} \mathbf{Q}) \, d\mathbf{r} - V_N \right]$ where $F$ is defined by Eq. \eqref{eqn:freeE}. The gradient flow equations are then
\begin{equation} \label{eqn:gradFlow}
    \frac{\partial \mathbf{Q}}{\partial t} = -\frac{\delta F}{\delta \mathbf{Q}} + \mu \mathbf{A}.
\end{equation}
Similar to Ref. \cite{wang18}, we calculate $\mu$ by taking a time derivative of the volume constraint, Eq. \eqref{eqn:VolConstraint}, so $\int_{\Omega} \mathbf{A} : \partial_t \mathbf{Q} = 0$. Substituting Eq. \eqref{eqn:gradFlow} for the time derivative of $\mathbf{Q}$ yields
\begin{equation}
    \mu = \frac{S_N^2}{18 V_{\Omega}} \int_{\Omega} \mathbf{A} : \frac{\delta F}{\delta \mathbf{Q}} \, d\mathbf{r}
\end{equation}
where $V_{\Omega}$ is the volume of the domain. Equation \eqref{eqn:gradFlow} is then solved by discretizing time and applying a semi-implicit convex splitting algorithm to solve the resulting equations as described above.\cite{schimming21} $\mu$ is recalculated at each time step, although it converges to be approximately constant in only a few iterations.

Using this volume constraint, we are able to fix the volume of both negative and positive tactoids. While this is not necessary for negative tactoids with a total topological charge, it is necessary for positive tactoids. Even if the temperature is set to be exactly at coexistence between the phases, positive tactoids tend to shrink due to the excess energy of the interface. If the temperature is set so that the nematic phase is lower energy than the isotropic phase, there will be an unstable tactoid size, below which the tactoid will shrink and above which the tactoid will grow indefinitely.\cite{chaikin95} The volume constraint allows us to stabilize positive tactoids of any size without coupling a phase field to the nematic field. Physically, the volume constraint may be thought of as associating a pressure to the fixed tactoid volume.

Finally, we note that this method is relatively simple for quasi two-dimensional systems where the director in the nematic phase is fixed in the $xy$ plane, because the matrix $\mathbf{A}$ is the same in both the fixed lab frame and the eigenframe of $\mathbf{Q}$. However, for fully three-dimensional nematics this is not the case, and $\mathbf{A}$ only has this simple form in the eigenframe, while $\mathbf{A}$ will depend on the director orientation in the fixed frame. Here we will only consider the one-dimensional and quasi two-dimensional cases, and avoid this complication. 

\section{Interfaces}
We first consider isotropic-nematic interfaces in one spatial dimension, and study the dependence of interface width and energy on the anchoring coefficient $L_4$ defined in Eq. \eqref{eqn:ElasticE}. Here the orientation of $\mathbf{n}$ is fixed, so there is no contribution of bulk elastic energy to the total energy, and the $L_3$ term has the same effect as the $L_4$ term. To obtain the configurations we first decompose $\mathbf{Q}$ into a convenient basis:
\begin{equation} \label{eqn:QBasis}
    \mathbf{Q} = \begin{pmatrix}
                \frac{2}{\sqrt{3}}q_1 & q_3 & q_4 \\
                q_3 & -\frac{1}{\sqrt{3}}q_1 + q_2 & q_5 \\
                q_4 & q_5 & -\frac{1}{\sqrt{3}}q_1 - q_2
                 \end{pmatrix}
\end{equation}
where $q_i$ are the degrees of freedom of the traceless, symmetric $\mathbf{Q}$. The energy is then written in terms of $q_i$ and the gradient flow equations are derived for each $q_i$. The discretized version of the gradient flow equations are solved using a semi-implicit convex-splitting method with 1000 finite elements and we set $\Delta t = 0.1$. We set the value of $\kappa / n T = 3.4049$ which is the value of exact coexistence between the nematic and isotropic phases\cite{schimming20}. Dirichlet conditions fix the left boundary to a nematic phase with eigenvalue $S_N = 0.4281$, the equilibrium value at this point of the phase diagram, and orientation of $\mathbf{n}$ given by $\phi_0$, the angle of $\mathbf{n}$ with respect to the interface. On the right boundary Dirichlet conditions fix the isotropic phase, $\mathbf{Q} = \mathbf{0}$. The gradient flow equations are iterated until the energy fails to change to within $10^{-6}$. We solve the governing equations with a range of 11 values of $\phi_0$ equally spaced from $0 \leq \phi_0 \leq \pi /2$ and the parameter $L_4$ up to $L_4 = 4$.

\begin{figure}[h]
\centering
	\includegraphics[width = 3.3in]{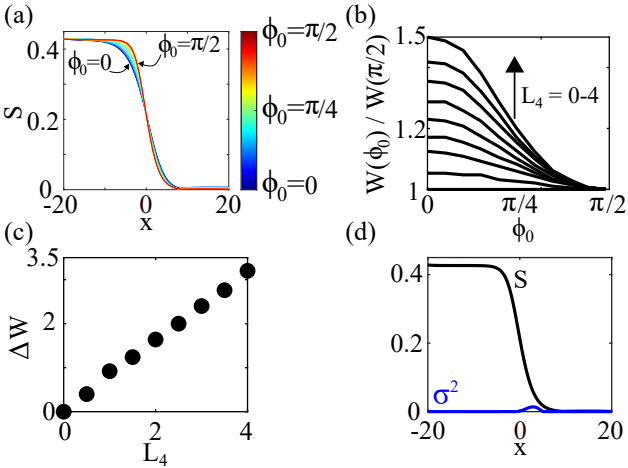}
	\caption{Nematic-isotropic interface for varying anchoring angle, $0 < \phi_0 < \pi/2$. (a) Plots of nematic order paramter $S$ versus position for $L_4 = 4$ and varying $\phi_0$, represented by the different colors. As the anchoring angle goes from homeotropic to tangential, the width of the interface decreases. (b) Interface width as a function of anchoring angle, normalized by the width at $\phi_0 = \pi /2$. Each curve represents a different value of $L_4$, with $L_4$ increasing with width disparity. (c) Width difference $\Delta W = W(\phi_0 =0) - W(\phi_0 = \pi/2)$ versus $L_4$. The linear relationship indicates that anchoring energy is proportional to $L_4$. (d) Nematic order, $S$, and biaxiality parameter, $\sigma^2$, across the interface for $L_4 = 4$ and $\phi_0 = \pi/2$. There is some biaxiality at the interface, though not large in magnitude.}
	\label{fig:Interface}
\end{figure}

Figure \ref{fig:Interface}a shows $S$ as a function of position for interfaces with various $\phi_0$ and $L_4 = 4$. Note that the interface width decreases and the interface itself becomes more asymmetric as $\phi_0$ is increased. The interface width becomes smaller with larger $\phi_0$ because the surface anchoring favors a parallel orientation. We expect the interfacial width to be proportional to the excess surface energy, since smaller gradients (and hence, larger widths) will cost less energy. In Fig. \ref{fig:Interface}b we show the width of the interface as a function of $\phi_0$ normalized to the width at $\phi_0 = \pi / 2$, for increasing $L_4$ values. Note that for $L_4 = 0$ the width does not depend on $\phi_0$. We show the normalized width since the $L_4$ term also changes the overall surface tension and there is no contribution from anchoring at $\phi_0 = \pi /2$ so the normalization captures the effect of surface anchoring. To calculate the width we define the points $x_1$ and $x_2$ where $S = 0.9S_N$ and $S = 0.1S_N$ respectively, then define the width by $W = x_2 - x_1$. Figure \ref{fig:Interface} shows the width differences $\Delta W = W(\phi_0 = 0) - W(\phi_0 = \pi/2)$ for various values of $L_4$. As expected, this width difference scales linearly with $L_4$. This is a good indication that it is reasonable to decompose $f_e(\mathbf{Q},\mathbf{\nabla}\mathbf{Q})$ into separate contributions to surface tension, surface anchoring, and bulk elasticity, as we have discussed in the previous section.

Finally, we examine the biaxiality of the interface for large anchoring. Figure \ref{fig:Interface}d shows the biaxiality parameter $\sigma^2 = 1 - 6 \Tr[\mathbf{Q}^3]^2 / \Tr[\mathbf{Q}^2]^3$ across the interface for $L_4 = 4$ and $\phi_0 = \pi /2$. As required by symmetry, $\sigma^2 = 0$ for $\phi_0 = 0$. However, we do see small amounts of biaxiality for $\phi_0 = \pi /2$. The largest region of biaxiality appears on the isotropic side of the interface, which likely causes the asymmetry in interface shape seen in Fig. \ref{fig:Interface}a. Unfortunately, the degree of biaxiality is not large enough here to qualitatively see in plots of the orientational probability distribution unlike what one sees around disclinations.\cite{schimming21} Because our interaction energy promotes uniaxial alignment (Eq. \eqref{eqn:BulkE}) the appearance of biaxiality is purely entropic. Our entropy is computed numerically in these computations, hence it is difficult to compare quantitatively to the Landau-de Gennes free energy, where the cubic order term promotes biaxiality.\cite{lyuksyutov78} We note that the angle dependent width and structure of the biaxiality of the interface is similar to previous theoretical and computational explorations of the isotropic-nematic interface.\cite{popanita97,koch99,wolfsheimer06,everts16}

\section{Negative Tactoids}
For the rest of the paper we will discuss computations in quasi two-dimensions. The qualifier ``quasi'' here refers to the fact that $\mathbf{Q}$ is a fully three-dimensional order parameter that is only allowed to vary in two spatial dimensions, i.e. $\mathbf{Q} = \mathbf{Q}(x,y)$. This is motivated by experiments performed in thin films between treated glass plates.\cite{kim13,zhou17} Additionally, a fully three-dimensional $\mathbf{Q}$ is required to model coexistence \cite{deGennes75}. Experiments performed on lyotropic chromonic liquid crystals have seen large negative tactoids with pronounced cusps. These cusps are actually surface defects, known as ``boojums,'' and are seen in positive tactoids as well, as we discuss in the next section.

\begin{figure}[h]
\centering
	\includegraphics[width = 3in]{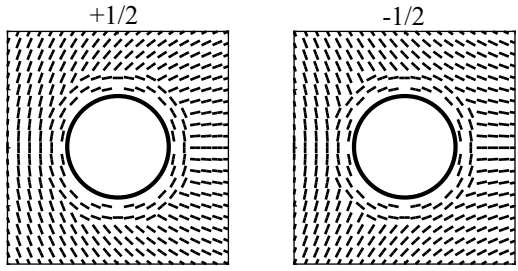}
	\caption{Initial condition of the director in the calculations involving negative tactoids with $\pm 1/2$ overall charge. This initial condition promotes the formation of boojums on the surface of the tactoid.}
	\label{fig:NegativeInitial}
\end{figure}

Negative tactoids typically result from melting disclinations, and as such they usually have a topological charge associated with them even though they are not technically defects themselves. Here we obtain negative tactoids with $\pm 1/2$ overall charge. Because of the topological charge, the volume constraint introduced in section 2 is not strictly needed, however we still employ it to obtain tactoids of similar size to experiments. It is difficult to observe spontaneously generated surface boojums in the mean-field calculation we use due to their high energy cost of nucleation. In order to study configurations that include them, we initialize the system with the director fields seen in Fig. \ref{fig:NegativeInitial}. This creates defects away from the tactoid, which eventually coalesce with the tactoid and form boojums on its surface.

\begin{figure}[h]
\centering
	\includegraphics[width = 3.4in]{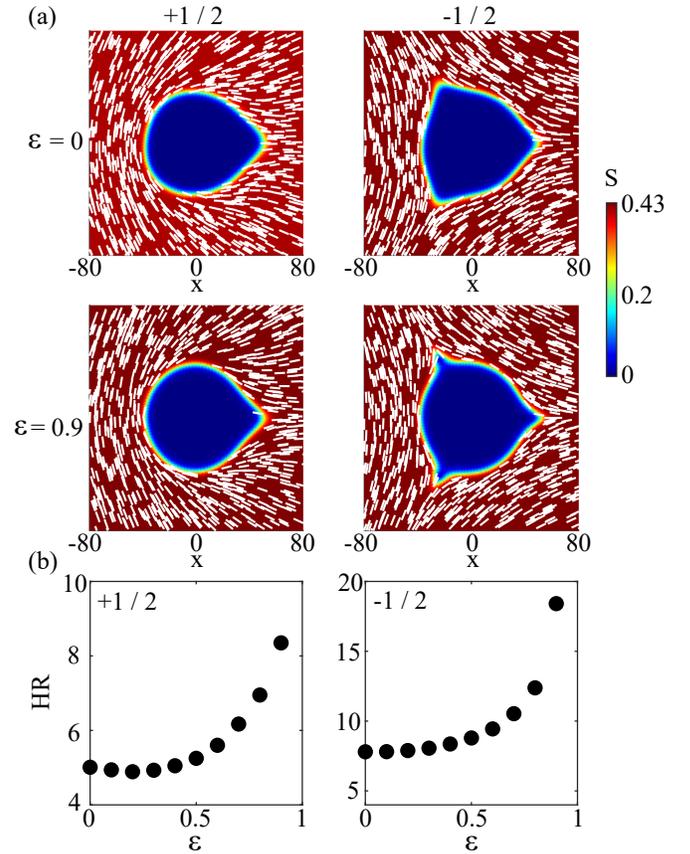}
	\caption{Numerically computed equilibrium structure of negative tactoids with varying elastic anisotropy. (a) Configurations for tactoids with topological character $+1 / 2$ (left) and $-1 / 2$ (right). The presented configurations vary from no elastic anisotropy, $\varepsilon = 0$, to large elastic anisotropy $\varepsilon = 0.9$. (b) Boojum curvature, $H$, times average tactoid radius, $R$, as a function of elastic anisotropy, $\varepsilon$. We find that the curvature in both $\pm 1 / 2$ negative tactoids increases dramatically with large elastic anisotropy.}
	\label{fig:NegativeTactoids}
\end{figure}

Our numerical solution is obtained in a $150 \times 150$ body-centered square mesh with $44701$ vertices. We solve the finite element discretized time-dependent gradient flow equations with $\Delta t = 0.1$ until the energy fails to change to within $10^{-6}$ of its current value. As with the one-dimensional interface, the parameter $\kappa / n T = 3.4049$. The outer boundaries of the domain are given Neumann boundary conditions. We fix the area of the tactoids to $40^2 \pi$ in dimensionless units. In all calculations, we maintain $L_3 + L_4 = 5$ so that the contributions to surface tension and surface anchoring from the elastic energy remain constant. Figure \ref{fig:NegativeTactoids}a shows negative tactoid configurations for no elastic anisotropy, $\varepsilon = (K_{33} - K_{11}) / (K_{33} + K_{11}) = 0$, and large elastic anisotropy, $\varepsilon = 0.9$. The elastic anisotropy parameter $\varepsilon$, defined in terms of the Frank-Oseen bend and splay constants, is a convenient measure of elastic anisotropy since it goes to $1$ ($-1$) in the limit $K_{11} = 0$ ($K_{33} = 0$). Qualitatively, we find that the cusp-like features created by the boojums become more prominent as $\varepsilon$ is increased. The large $\varepsilon$ plots are qualitatively similar to experiments, particularly those seen in Ref. \cite{zhou17}. We note that in our calculations we have chosen to set the area of tactoids so that the ratio of interfacial width to tactoid radius is approximately equal to those seen in the experiments.

In order to quantify the sharpening of the cusps, Fig. \ref{fig:NegativeTactoids}b shows boojum curvature times average tactoid radius, $H R$, versus the elastic anisotropy parameter. Boojum curvature is computed by fitting a parabola to the points on the interface close to the boojum; the curvature is then extracted from the resulting fit. We note that the curvature of the boojums increases rapidly for increasing $\varepsilon > 0.5$. Therefore we infer that the appearance of cusps in tactoids is linked to the elastic anisotropy of the material. This can be understood intuitively as a balance between anchoring energy---promoting tangential alignment---and bulk elastic energy where splay dominates at the site of the boojums. If the splay constant is reduced, as is the case when $\varepsilon > 0$, the tactoid may incur more splay distortion to maintain its tangential anchoring at the boojum. Unfortunately, there is not a quantitative analysis of the cusp-like boojums of negative tactoids in the experimental literature that we are aware of to compare to.

\section{Positive Tactoids}

\begin{figure}[h]
\centering
	\includegraphics[width = 3.2in]{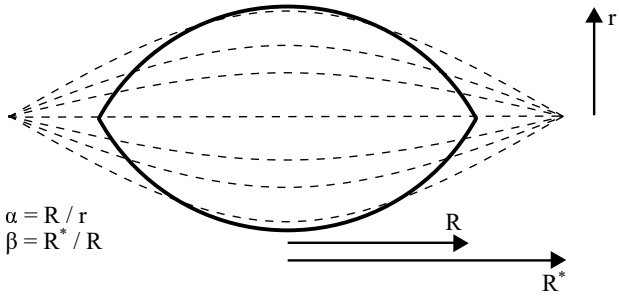}
	\caption{Sketch of a tactoid. Solid lines represent the interface while dashed lines represent the director field which converges on the location of virtual defects.}
	\label{fig:TactoidSketch}
\end{figure}

We now discuss the equilibrium properties of positive tactoids where a domain of nematic phase is surrounded by the isotropic phase. We consider postive tactoids with director fields ranging from homogeneous, where the director is uniform, to bipolar, where the director field is characterized by defects at the ends of the domain. Configurations between these two limits can be succinctly characterized by the location of ``virtual'' defects given by extending the director into the isotropic phase a distance $R^{*}$ from the center of the tactoid. In addition to the director configuration, we parameterize the boundary shape obtained numerically by two arcs of a circle, giving rise to a long axis $R$ and short axis $r$. This is typically the assumed shape in other numerical studies and is seen in tactoids in experiments.\cite{prinsen03,prinsen04,kim13,paparini21} Hence, there are two dimensionless parameters that we study: the aspect ratio, $\alpha \equiv R/r$, and the ratio of the distance of the virtual defect location to the long axis length of the tactoid, $\beta \equiv R^{*} / R$. A sketch of a tactoid with these parameters indicated is shown in Fig. \ref{fig:TactoidSketch}. The minimum aspect ratio is $\alpha = 1$ which represents a circular domain. $\beta$ ranges from $1$ to $\infty$ where $1$ represents a perfectly bipolar director configuration with boojums at the ends, and $\infty$ represents a perfectly homogeneous configuration. We note that roughly if $\beta \geq 3$, the director configuration becomes nearly indistinguishable from the homogeneous case.

To investigate the effect of anisotropic elasticity on these parameters, we computationally determine equilibrium tactoid shapes in a similar manner as laid out in Sec. 4 for the case of negative tactoids. We fix the parameters $L_3 + L_4$ and $\varepsilon$, as well as the area of the positive tactoids (to $120^2 \pi$ in dimensionless units), and initialize them with varying $\{\alpha,\beta\}$ such that $\alpha \in \{1,\,1.2,\,1.4,\,1.6,\,1.8,\,2\}$ and $\beta \in \{1,\,1.05,\,1.1,\,1.2,\,1.5,\,2,\,\infty\}$. We then compute the energy of the configuration after it fails to change to within $10^{-6}$ of its current value. The energy landscape is quite flat, hence the director fields and interface shapes do not change much from their initialized configuration.

\begin{figure}[h]
\centering
	\includegraphics[width = 3in]{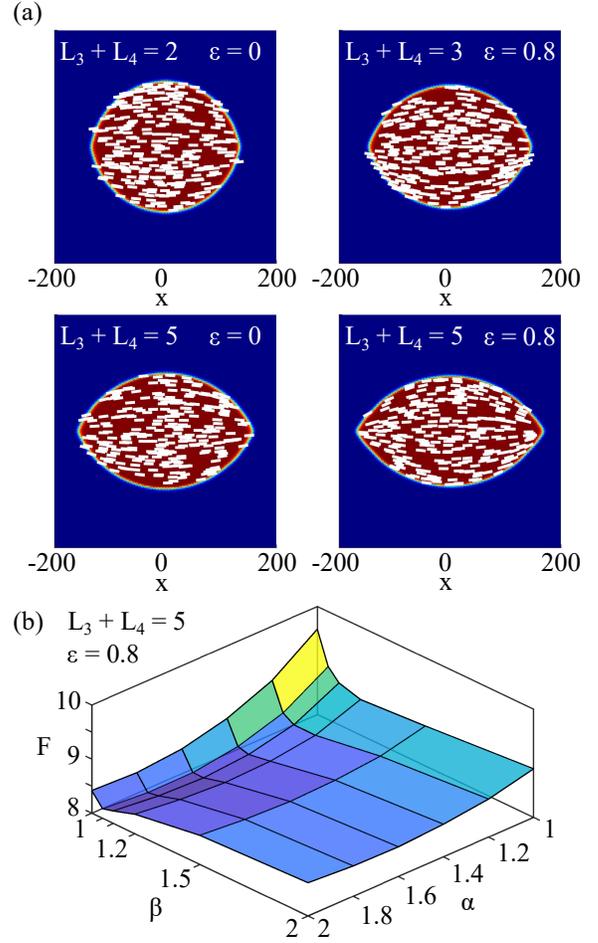}
	\caption{(a) Minimum energy tactoid configurations for various anchoring strength $L_3 + L_4$ and elastic anisotropy $\varepsilon$. As anchoring strength increases, the optimal aspect ratio increases. As elastic anisotropy increases, the director configuration becomes more bipolar. (b) Plot of the free energy $F$ as a function of aspect ratio $\alpha$ and bipolarity measure $\beta$ for $L_3 + L_4 = 5$ and $\varepsilon = 0.8$. The energy landscape is relatively flat with the highest energy configuration being the circular and bipolar configuration $\{\alpha,\,\beta\} = \{1,\,1\}$.}
	\label{fig:PosTacPlots}
\end{figure}

Figure \ref{fig:PosTacPlots}a shows the computed minimum energy configurations for various values of anchoring strength, measured by $L_3 + L_4$, and elastic anisotropy, measured by $\varepsilon$. Figure \ref{fig:PosTacPlots}b shows an example energy landscape for $L_3 + L_4 = 5$ and $\varepsilon = 0.8$, as a function of $\alpha$ and $\beta$. We note that the free energy landscape is relatively flat and the highest energy configuration is typically the circular, bipolar case with $\{\alpha,\,\beta\} = \{1,\,1\}$. Table \ref{table:minETacs} shows the parameter sets, $\{\alpha,\,\beta\}$, that yield a minimum in the free energy for each set of $L_3 + L_4$ and $\varepsilon$ studied. The minimum energy configurations show that as elastic anisotropy is increased, tactoids become more bipolar, i.e. $\beta$ decreases. Additionally, an increase in anchoring strength is associated with an increase in aspect ratio $\alpha$. For large anchoring, elastic anisotropy also tends to increase the aspect ratio of the tactoid. It has been hypothesized in previous studies of tactoids with a $\mathbf{Q}$-tensor approach that considering cubic order terms in $\mathbf{Q}$ in the elastic energy may be required to observe larger aspect ratios and bipolarity.\cite{everts16} Here we have shown that this is indeed the case.

\begin{table*}
\small
  \caption{\ Parameters $\{\alpha,\,\beta \}$ of the minimum energy positive tactoid with anchoring strength measured by $L_3 + L_4$ and elastic anisotropy $\varepsilon$.}
  \label{table:minETacs}
  \begin{tabular*}{\textwidth}{@{\extracolsep{\fill}}llllll}
    \hline
    $L_3 + L_4$ & $\varepsilon = 0$ & $\varepsilon = 0.2$ & $\varepsilon = 0.4$ & $\varepsilon = 0.6$ & $\varepsilon = 0.8$\\
    \hline
    $0$ & $\{1,\, \infty \}$ & $\{1,\, \infty \}$ & $\{1,\, \infty \}$ & $\{1,\, \infty \}$ & $\{1,\, \infty \}$ \\
    $1$ & $\{1.2,\, 2 \}$ & $\{1.2,\, 2 \}$ & $\{1.2,\, 2 \}$ & $\{1.2,\, 1.5 \}$ & $\{1.2,\, 1.5 \}$ \\
    $2$ & $\{1.2,\, 1.5 \}$ & $\{1.2,\, 1.5 \}$ & $\{1.2,\, 1.5 \}$ & $\{1.2,\, 1.5 \}$ & $\{1.2,\, 1.5 \}$ \\
    $3$ & $\{1.4,\, 1.5 \}$ &  $\{1.4,\, 1.2 \}$ & $\{1.4,\, 1.2 \}$ & $\{1.4,\, 1.2 \}$ & $\{1.4,\, 1.2 \}$ \\
    $4$ & $\{1.6,\, 1.2 \}$ & $\{1.6,\, 1.2 \}$ & $\{1.6,\, 1.1 \}$ & $\{1.6,\, 1.1 \}$ & $\{1.6,\, 1.1 \}$ \\
    $5$ & $\{1.6,\, 1.2 \}$ & $\{1.6,\, 1.2 \}$ & $\{1.8,\, 1.1 \}$ & $\{1.8,\, 1.1 \}$ & $\{1.8,\, 1.05 \}$ \\
    \hline
  \end{tabular*}
\end{table*}

We finally note that while the shapes of the positive tactoids are similar to those seen in experiments, the minimum energy configurations are slightly different. Tactoids appearing in Ref. \cite{kim13} are more bipolar and have a smaller aspect ratio than the minimum energy configurations in our study. Based on the theoretical work in Ref. \cite{vanderschoot12} this can be explained by a larger surface tension to elastic constant ratio in the experiments. This disparity highlights a shortcoming in the $\mathbf{Q}$-tensor approach; namely, there are no elastic energy terms that only contribute to surface tension, thus making a study of the separate effect of surface tension difficult. 

\section{Tactoid Growth}
Finally, we explore the effect of the elastic constants $L_3$ and $L_4$ on nonequilibrium tactoid growth. To do this, we remove the volume constraint of the previous sections, and lower the effective temperature so as to simulate a quench into the nematic phase. We set $\kappa / n T = 3.6$, which is set so that the isotropic phase is still metastable, but the nematic phase is heavily favored by the bulk free energy. We initialize the system with a circular, homogeneous configuration of radius $R = 20$, though we mention that initializing with a bipolar configuration does not change the long-time results that we present here. We then numerically compute $500$ iterations of growth with $\Delta t = 0.2$.

\begin{figure*}
	\includegraphics[width = 7in]{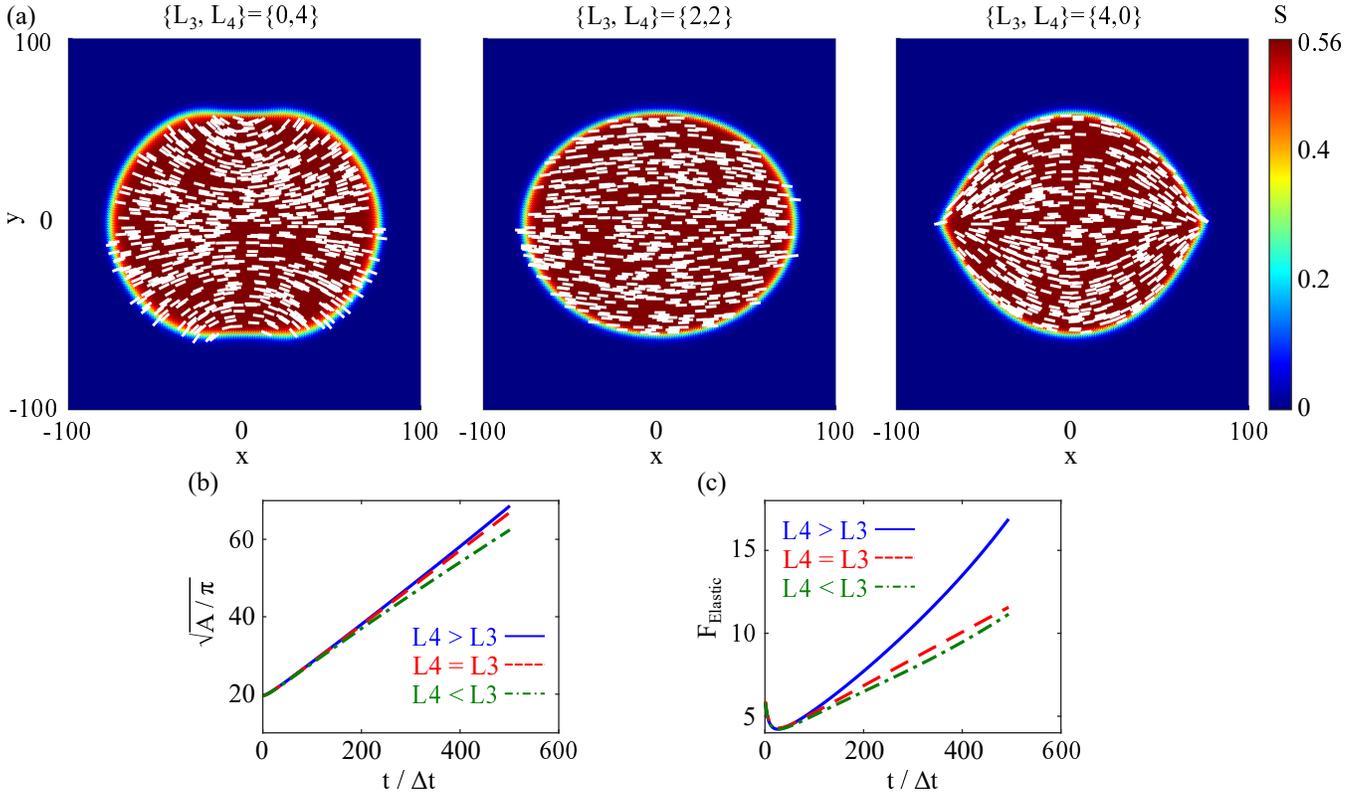}
	\caption{Numerical computation of tactoid growth from a homogeneous circular initial condition for parameters $\{L_3,\,L_4\} = \{0,\,4\}$, $\{2,\,2\}$, $\{4,\,0\}$. (a) Tactoid configurations at iteration number $t / \Delta t = 500$. The tactoid with $L_4 < L_3$ develops defects at its ends and takes on a bipolar director configuration, while the tactoid with $L_4 > L_3$ grows homeotropically and eventually develops defects at its top and bottom ends. The tactoid with $L_4 = L_3$ grows anisotropic in shape but does not develop heterogeneity in the director configuration. (b) Plots of the square root of the area, $A$, versus iteration number $t / \Delta t$. The area of the homeotropic tactoid grows fastest due to a larger interface velocity for homeotropic anchoring. (c) Plots of the elastic free energy, Eq. \eqref{eqn:ElasticE}, versus iteration number. The bipolar tactoid grows to have the lowest elastic free energy of all configurations.}
	\label{fig:TactoidGrowth}
\end{figure*}

Figure \ref{fig:TactoidGrowth}a shows results for three particular cases: $\{L_3,\,L_4\} = \{0,\,4\}$, $\{2,\,2\}$, $\{4,\,0\}$. In all three cases, the area of the tactoids grows like $A \sim t^2$ (Fig. \ref{fig:TactoidGrowth}b) which is expected for two dimensional domain growth driven by phase free energy difference.\cite{chaikin95} However, the director configurations are markedly different in all three cases. We find that if $L_3 > L_4$, the director develops a bipolar configuration with two boojums at the ends of the tactoid, similar to the shapes studied in the previous section, and similar to shapes seen in experiments.\cite{kim13,jamali15} This configuration minimizes the anchoring energy by eventually orienting the director tangential to the interface. On the other hand, if $L_4 > L_3$ the director tends to homeotropic alignment, eventually forming defects on the top and bottom poles of the tactoid. This configuration does \textit{not} minimize the elastic energy associated with anchoring, nor does it minimize the Frank elastic energy for bulk director variations. This is seen in Fig. \ref{fig:TactoidGrowth}c which shows the elastic energy as a function of iteration number, $t / \Delta t$. When $L_3 = L_4$ the effect is effectively cancelled out and the director remains homogeneous, though the tactoid still grows anisotropically. The elastic energy of this configuration is only slightly larger than that of the bipolar configuration since now the anchoring energy is increased but there is no Frank elastic energy in the bulk.

The results shown in Fig. \ref{fig:TactoidGrowth} are similar to numerical results from Ref. \cite{wincure06}. There, the effect of a different anisotropic elastic term, quadratic in $\mathbf{Q}$ (referred to as the $L_2$ term) was studied alongside the effect of the $L_3$ term. Here, since both the $L_3$ and $L_4$ terms are cubic in $\mathbf{Q}$, their effects on the director kinetics at the interface can be cancelled out by having equal coefficients. This leads to a simple explanation for the phenomenon: there is a term contained in $\delta F_3 / \delta \mathbf{Q}$ and in $\delta F_4 / \delta \mathbf{Q}$ that is exactly equal and of opposite sign, and this term must contribute to the tendency of the director to become homeotropic or tangential during tactoid growth. Here $F_3$ and $F_4$ refer to the elastic energies from the $L_3$ and $L_4$ terms respectively. Computation of these functional derivatives yields
\begin{align}
    \frac{\delta F_3}{\delta Q_{k \ell}} &= \partial_k Q_{ij} \partial_{\ell} Q_{ij} - 2 \partial_i Q_{k \ell}\partial_j Q_{ij} - 2 Q_{ij}\partial_k\partial_{\ell}Q_{ij} \label{eqn:F3}\\
    \frac{\delta F_4}{\delta Q_{k \ell}} &= -\partial_k Q_{ij} \partial_{\ell} Q_{ij} - Q_{ij} \partial_{k}\partial_{\ell} Q_{ij} - Q_{k \ell}\partial_i \partial_j Q_{ij}. \label{eqn:F4}
\end{align}

The first term in Eqs. \eqref{eqn:F3} and \eqref{eqn:F4} are equal and of opposite sign, therefore we will investigate the effect of this term on the eigenvectors of $\mathbf{Q}$, which represent the director. To do this, we parameterize $\mathbf{Q}$ as fully biaxial: $\mathbf{Q} = S\left[\mathbf{n} \otimes \mathbf{n} - 1/3\mathbf{I}\right] + P\left[\mathbf{m}\otimes\mathbf{m} - \bm{\ell}\otimes\bm{\ell}\right]$ where $\mathbf{n}$, $\mathbf{m}$, and $\bm{\ell}$ are mutually orthogonal and normalized, and represent the eigenvectors of $\mathbf{Q}$ while $S$ and $P$ parameterize its eigenvalues. Further, since we are working in two dimensions, we can write $\mathbf{n} = \left(\cos\varphi,\,\sin\varphi,\,0\right)$, $\mathbf{m} = \left(-\sin\varphi,\,\cos\varphi,\,0\right)$, and $\bm{\ell} = \left(0,\,0,\,1\right)$ where $\varphi$ is the angle of the director with respect to the $x$ axis. We use this parameterization to compute 
\begin{equation} \label{eqn:mdtQn}
 \mathbf{m} \cdot \partial_t \mathbf{Q} \cdot \mathbf{n} = S \partial_t \varphi.    
\end{equation}
Note that this projection picks out the time dependence of the director. We now consider the first term in Eqs. \eqref{eqn:F3} and \eqref{eqn:F4}. Since $\partial_t \mathbf{Q} \propto \delta F / \delta \mathbf{Q}$, we compute the same projection on $\partial_t\mathbf{Q}$ above on the first term in Eqs. \eqref{eqn:F3} and \eqref{eqn:F4}:
\begin{multline} \label{eqn:mdeltaQn}
    m_k \partial_k Q_{ij} \partial_{\ell} Q_{ij} n_{\ell} = \frac{2}{3}m_k \partial_k S n_{\ell} \partial_{\ell} S + 2m_k \partial_k P n_{\ell} \partial_{\ell} P \\
    + 2(S - P)^2 m_k \partial_k \varphi n_{\ell} \partial_{\ell} \varphi.
\end{multline}
We note that, at the interface, the second and third terms of the above equation should be small compared to the first, and that the gradient of $S$ is in the direction of the interface normal. Thus, comparing Eqs. \eqref{eqn:mdtQn} and \eqref{eqn:mdeltaQn} gives $\partial_t\varphi \propto (L_4 - L_3)\sin2\left(\theta - \varphi\right)$ where $\theta$ is the angle of the interface normal with respect to the $x$ axis. This contribution vanishes when $L_3 = L_4$, and when the director is tangential or perpendicular to the interface, but depending on the relative elastic coefficients, it either drives the director to be one or the other.

Since this effect on the kinetics of director orientation at the interface deriving from the $L_3$ and $L_4$ elastic terms is equal and opposite, we conclude that this feature of tactoid growth is \textit{not} due to elastic anisotropy. While the bulk director field may evolve to minimize the relevant terms in the Frank-Oseen energy, the tendency for boojums to form in the $\mathbf{Q}$-tensor elastic energy is generated by different terms. We therefore emphasize that this process is of kinetic origin, and that even if $L_4 > L_3$ the configuration with minimum energy is still one with the director tangential to the interface. Of course, the full free energy, that is, bulk plus elastic free energy, is still decreasing at each time step since the nematic domain is growing. Further, we note that the bipolar configuration seen in Fig. \ref{fig:TactoidGrowth} does not constitute an energy minimum since this grows to have $\beta \approx 1$ while the minimum energy configuration for fixed volume and similar $\varepsilon$ found in Section 5 has $\beta \approx 1.1$.

\begin{figure}[h]
	\includegraphics[width = 3in]{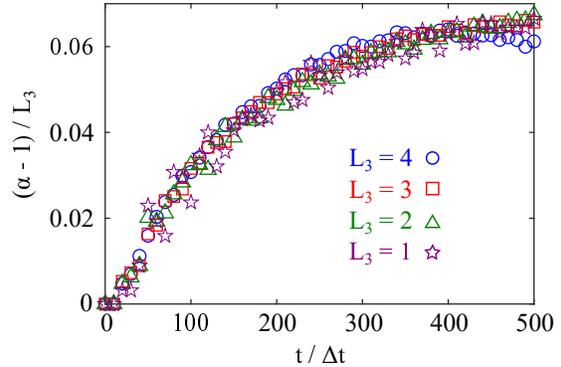}
	\caption{Tactoid aspect ratio, scaled as $(\alpha - 1) / L_3$, versus computation time step, $t / \Delta t$ for various values of $L_3$. The data collapse indicates that both the growth and steady state aspect ratio is linearly proportional to the value of $L_3$.}
	\label{fig:AspectRatios}
\end{figure}

We finish this section by exploring the configurations that develop at long times with variable $L_3$. If $L_3 > 0$, the tactoid will develop a bipolar director configuration, leading to an anisotropic morphology with aspect ratio $\alpha > 1$. The aspect ratio continues to grow until boojums form, at which point the aspect ratio tends to stay constant since the director is tangent to the interface everywhere except the endpoints. Figure \ref{fig:AspectRatios} shows $(1 - \alpha) / L_3$ as a function of time step. Note that the data collapse for various $L_3$ indicates that the growth and saturation of aspect ratio in tactoids is linearly proportional to $L_3$. We emphasize that this does not reflect tactoid growth slowing, since Fig. \ref{fig:TactoidGrowth}b shows that $\sqrt{A} \sim t$ for all times. Instead, the figure shows that the aspect ratio at long times is determined kinetically by the value of $L_3$.

\section{Conclusion}
In summary, we have presented a numerical study of nematic-isotropic interfaces, and of negative and positive tactoids by using a $\mathbf{Q}$-tensor representation of nematic order. A singular Maier-Saupe bulk energy and cubic order anisotropic elastic energy have been included in the free energy. We have further developed and demonstrated a numerical technique involving Lagrange multipliers to constrain the volume of a nematic phase which allows for the study of equilibrium morphologies exactly at nematic-isotropic coexistence.

The theory has allowed us to probe the effect of anisotropic elasticity on interfacial thickness, anchoring energy, and biaxiality at the interface. Larger anisotropy leads to a larger energy difference between homeotropic and tangential anchoring and thus to stronger tangential anchoring. In two dimensional systems, anisotropic Frank-Oseen elasticity leads to sharp boojums in negative tactoids, which are also seen in experiments in lyotropic chromonic liquid crystals. On the other hand, sharp boojums are not formed in positive tactoids, and instead the elastic energy favors a director configuration between bipolar and homogeneous, with larger anisotropy favoring more bipolarity. Finally, the kinetic growth of positive tactoids is shown to be very sensitive to the anisotropic term included in the energy, with bipolar, anisotropic tactoids generated by the $L_3$ cubic elastic term in Eq. \eqref{eqn:ElasticE}.

Our results should prove useful to future computational and experimental studies of tactoids in nematic liquid crystals. The primary advantage of the $\mathbf{Q}$-tensor framework is the ability to resolve defects such as boojums, while also capturing the smooth transition in order parameter from the nematic to isotropic phase. The volume constraint introduced allows for the volume of the tactoid to be a control parameter, and allows for two phase interfaces without additional conserved fields. For future studies, it will be interesting to explore the possibility of a full three-dimensional volume constraint. Additionally, the work presented here on tactoid growth assumed relaxational dynamics of the order parameter with no hydrodynamic transport. Unlike chromonics, a density contrast between the isotropic and nematic phases is generally observed in lyotropic liquid crystals. Such density contrast will induce mass transport during tactoid growth, which will need to be examined.\cite{re:vitral21} It would also be of interest to study the coalescence of tactoids and the resulting nucleation of defects, and the effect of anisotropy on this process.


\section*{Conflicts of interest}
There are no conflicts to declare.

\section*{Acknowledgements}
This research has been supported by the National Science Foundation under Grant No. DMR-1838977, and by the Minnesota Supercomputing Institute.



\balance

\renewcommand\refname{References}

\bibliography{LC} 

\providecommand*{\mcitethebibliography}{\thebibliography}
\csname @ifundefined\endcsname{endmcitethebibliography}
{\let\endmcitethebibliography\endthebibliography}{}
\begin{mcitethebibliography}{44}
\providecommand*{\natexlab}[1]{#1}
\providecommand*{\mciteSetBstSublistMode}[1]{}
\providecommand*{\mciteSetBstMaxWidthForm}[2]{}
\providecommand*{\mciteBstWouldAddEndPuncttrue}
  {\def\EndOfBibitem{\unskip.}}
\providecommand*{\mciteBstWouldAddEndPunctfalse}
  {\let\EndOfBibitem\relax}
\providecommand*{\mciteSetBstMidEndSepPunct}[3]{}
\providecommand*{\mciteSetBstSublistLabelBeginEnd}[3]{}
\providecommand*{\EndOfBibitem}{}
\mciteSetBstSublistMode{f}
\mciteSetBstMaxWidthForm{subitem}
{(\emph{\alph{mcitesubitemcount}})}
\mciteSetBstSublistLabelBeginEnd{\mcitemaxwidthsubitemform\space}
{\relax}{\relax}

\bibitem[de~Gennes(1975)]{deGennes75}
P.~G. de~Gennes, \emph{The Physics of Liquid Crystals}, Oxford University
  Press, 1975\relax
\mciteBstWouldAddEndPuncttrue
\mciteSetBstMidEndSepPunct{\mcitedefaultmidpunct}
{\mcitedefaultendpunct}{\mcitedefaultseppunct}\relax
\EndOfBibitem
\bibitem[Prinsen and van~der Schoot(2003)]{prinsen03}
P.~Prinsen and P.~van~der Schoot, \emph{Phys Rev. E}, 2003, \textbf{68},
  021701\relax
\mciteBstWouldAddEndPuncttrue
\mciteSetBstMidEndSepPunct{\mcitedefaultmidpunct}
{\mcitedefaultendpunct}{\mcitedefaultseppunct}\relax
\EndOfBibitem
\bibitem[Prinsen and van~der Schoot(2004)]{prinsen04}
P.~Prinsen and P.~van~der Schoot, \emph{Eur. Phys. J. E}, 2004, \textbf{13},
  35--41\relax
\mciteBstWouldAddEndPuncttrue
\mciteSetBstMidEndSepPunct{\mcitedefaultmidpunct}
{\mcitedefaultendpunct}{\mcitedefaultseppunct}\relax
\EndOfBibitem
\bibitem[Prinsen and van~der Schoot(2004)]{prinsen04b}
P.~Prinsen and P.~van~der Schoot, \emph{J. Phys.: Condens. Matter}, 2004,
  \textbf{16}, 8835\relax
\mciteBstWouldAddEndPuncttrue
\mciteSetBstMidEndSepPunct{\mcitedefaultmidpunct}
{\mcitedefaultendpunct}{\mcitedefaultseppunct}\relax
\EndOfBibitem
\bibitem[Wincure and Rey(2006)]{wincure06}
B.~Wincure and A.~D. Rey, \emph{J. Chem. Phys.}, 2006, \textbf{124},
  244902\relax
\mciteBstWouldAddEndPuncttrue
\mciteSetBstMidEndSepPunct{\mcitedefaultmidpunct}
{\mcitedefaultendpunct}{\mcitedefaultseppunct}\relax
\EndOfBibitem
\bibitem[van Bijnen \emph{et~al.}(2012)van Bijnen, Otten, and van~der
  Schoot]{vanderschoot12}
R.~M.~W. van Bijnen, R.~H.~J. Otten and P.~van~der Schoot, \emph{Phys. Rev. E},
  2012, \textbf{86}, 051703\relax
\mciteBstWouldAddEndPuncttrue
\mciteSetBstMidEndSepPunct{\mcitedefaultmidpunct}
{\mcitedefaultendpunct}{\mcitedefaultseppunct}\relax
\EndOfBibitem
\bibitem[Zhang \emph{et~al.}(2018)Zhang, Acharya, Walkington, and
  Lavrentovich]{zhang18}
C.~Zhang, A.~Acharya, N.~J. Walkington and O.~D. Lavrentovich, \emph{Liq.
  Cryst.}, 2018, \textbf{45}, 1084\relax
\mciteBstWouldAddEndPuncttrue
\mciteSetBstMidEndSepPunct{\mcitedefaultmidpunct}
{\mcitedefaultendpunct}{\mcitedefaultseppunct}\relax
\EndOfBibitem
\bibitem[Golovaty \emph{et~al.}(2020)Golovaty, Kim, Lavrentovich, Novack, and
  Sternberg]{golovaty20}
D.~Golovaty, Y.-K. Kim, O.~D. Lavrentovich, M.~Novack and P.~Sternberg,
  \emph{Math. Model. Nat. Phenom.}, 2020, \textbf{15}, 8\relax
\mciteBstWouldAddEndPuncttrue
\mciteSetBstMidEndSepPunct{\mcitedefaultmidpunct}
{\mcitedefaultendpunct}{\mcitedefaultseppunct}\relax
\EndOfBibitem
\bibitem[Paparini and Virga(2021)]{paparini21}
S.~Paparini and E.~G. Virga, \emph{Phys. Rev. E}, 2021, \textbf{103},
  022707\relax
\mciteBstWouldAddEndPuncttrue
\mciteSetBstMidEndSepPunct{\mcitedefaultmidpunct}
{\mcitedefaultendpunct}{\mcitedefaultseppunct}\relax
\EndOfBibitem
\bibitem[Paparini and Virga(2021)]{paparini21b}
S.~Paparini and E.~G. Virga, \emph{J. Phys. Condens. Matter}, 2021,
  \textbf{33}, 495101\relax
\mciteBstWouldAddEndPuncttrue
\mciteSetBstMidEndSepPunct{\mcitedefaultmidpunct}
{\mcitedefaultendpunct}{\mcitedefaultseppunct}\relax
\EndOfBibitem
\bibitem[Lopez-Leon \emph{et~al.}(2011)Lopez-Leon, Koning, Devaiah, Vitelli,
  and Fernandez-Nieves]{lopez11}
T.~Lopez-Leon, V.~Koning, K.~B.~S. Devaiah, V.~Vitelli and A.~Fernandez-Nieves,
  \emph{Nat. Phys.}, 2011, \textbf{7}, 391--394\relax
\mciteBstWouldAddEndPuncttrue
\mciteSetBstMidEndSepPunct{\mcitedefaultmidpunct}
{\mcitedefaultendpunct}{\mcitedefaultseppunct}\relax
\EndOfBibitem
\bibitem[Lazo \emph{et~al.}(2014)Lazo, Peng, Xiang, Shiyanovskii, and
  Lavrentovich]{lazo14}
I.~Lazo, C.~Peng, J.~Xiang, S.~V. Shiyanovskii and O.~D. Lavrentovich,
  \emph{Nat. Commun.}, 2014, \textbf{5}, 5033\relax
\mciteBstWouldAddEndPuncttrue
\mciteSetBstMidEndSepPunct{\mcitedefaultmidpunct}
{\mcitedefaultendpunct}{\mcitedefaultseppunct}\relax
\EndOfBibitem
\bibitem[Peng \emph{et~al.}(2015)Peng, Guo, Conklin, Vi\~{n}als, Shiyanovskii,
  Wei, and Lavrentovich]{peng15}
C.~Peng, Y.~Guo, C.~Conklin, J.~Vi\~{n}als, S.~Shiyanovskii, Q.-H. Wei and
  O.~D. Lavrentovich, \emph{Phys. Rev. E}, 2015, \textbf{92}, 052502\relax
\mciteBstWouldAddEndPuncttrue
\mciteSetBstMidEndSepPunct{\mcitedefaultmidpunct}
{\mcitedefaultendpunct}{\mcitedefaultseppunct}\relax
\EndOfBibitem
\bibitem[Patteson \emph{et~al.}(2018)Patteson, Gopinath, and
  Arratia]{patteson18}
A.~E. Patteson, A.~Gopinath and P.~E. Arratia, \emph{Nat. Commun.}, 2018,
  \textbf{9}, 5373\relax
\mciteBstWouldAddEndPuncttrue
\mciteSetBstMidEndSepPunct{\mcitedefaultmidpunct}
{\mcitedefaultendpunct}{\mcitedefaultseppunct}\relax
\EndOfBibitem
\bibitem[Peng \emph{et~al.}(2018)Peng, Turiv, Guo, Wei, and
  Lavrentovich]{peng18}
C.~Peng, T.~Turiv, Y.~Guo, Q.-H. Wei and O.~D. Lavrentovich, \emph{Liq.
  Cryst.}, 2018, \textbf{45}, 1936\relax
\mciteBstWouldAddEndPuncttrue
\mciteSetBstMidEndSepPunct{\mcitedefaultmidpunct}
{\mcitedefaultendpunct}{\mcitedefaultseppunct}\relax
\EndOfBibitem
\bibitem[Babakhanova \emph{et~al.}(2018)Babakhanova, Turiv, Guo, Hendrikx, Wei,
  Schenning, Broer, and Lavrentovich]{baba18}
G.~Babakhanova, T.~Turiv, Y.~Guo, M.~Hendrikx, Q.-H. Wei, A.~P. Schenning,
  D.~J. Broer and O.~D. Lavrentovich, \emph{Nat. Commun.}, 2018, \textbf{9},
  456\relax
\mciteBstWouldAddEndPuncttrue
\mciteSetBstMidEndSepPunct{\mcitedefaultmidpunct}
{\mcitedefaultendpunct}{\mcitedefaultseppunct}\relax
\EndOfBibitem
\bibitem[Hokmabad \emph{et~al.}(2019)Hokmabad, Baldwin, Kr{\"u}ger, Bahr, and
  Maass]{hokmabad19}
B.~V. Hokmabad, K.~A. Baldwin, C.~Kr{\"u}ger, C.~Bahr and C.~C. Maass,
  \emph{Phys. Rev. Lett.}, 2019, \textbf{123}, 178003\relax
\mciteBstWouldAddEndPuncttrue
\mciteSetBstMidEndSepPunct{\mcitedefaultmidpunct}
{\mcitedefaultendpunct}{\mcitedefaultseppunct}\relax
\EndOfBibitem
\bibitem[Coelho \emph{et~al.}(2020)Coelho, Ara\'{u}jo, and Telo~da
  Gama]{coelho20}
R.~C.~V. Coelho, N.~A.~M. Ara\'{u}jo and M.~M. Telo~da Gama, \emph{Soft
  Matter}, 2020, \textbf{16}, 4256\relax
\mciteBstWouldAddEndPuncttrue
\mciteSetBstMidEndSepPunct{\mcitedefaultmidpunct}
{\mcitedefaultendpunct}{\mcitedefaultseppunct}\relax
\EndOfBibitem
\bibitem[Lee \emph{et~al.}(2020)Lee, Lee, Yoon, Hong, and Jang-Kun]{lee20}
B.~Lee, J.-S. Lee, H.-J. Yoon, S.-H. Hong and S.~Jang-Kun, \emph{Phys. Rev. E},
  2020, \textbf{101}, 012704\relax
\mciteBstWouldAddEndPuncttrue
\mciteSetBstMidEndSepPunct{\mcitedefaultmidpunct}
{\mcitedefaultendpunct}{\mcitedefaultseppunct}\relax
\EndOfBibitem
\bibitem[Ludwig \emph{et~al.}(2020)Ludwig, Weirich, Alster, Witten, Gardel,
  Dasbiswas, and Vaikuntanathan]{ludwig20}
N.~B. Ludwig, K.~Weirich, E.~Alster, T.~A. Witten, M.~L. Gardel, K.~Dasbiswas
  and S.~Vaikuntanathan, \emph{J. Chem. Phys.}, 2020, \textbf{152},
  084901\relax
\mciteBstWouldAddEndPuncttrue
\mciteSetBstMidEndSepPunct{\mcitedefaultmidpunct}
{\mcitedefaultendpunct}{\mcitedefaultseppunct}\relax
\EndOfBibitem
\bibitem[Everts \emph{et~al.}(2016)Everts, Punter, Samin, van~der Schoot, and
  van Roij]{everts16}
J.~C. Everts, M.~T. J. J.~M. Punter, S.~Samin, P.~van~der Schoot and R.~van
  Roij, \emph{J. Chem. Phys.}, 2016, \textbf{144}, 194901\relax
\mciteBstWouldAddEndPuncttrue
\mciteSetBstMidEndSepPunct{\mcitedefaultmidpunct}
{\mcitedefaultendpunct}{\mcitedefaultseppunct}\relax
\EndOfBibitem
\bibitem[Kaznacheev \emph{et~al.}(2002)Kaznacheev, Bogdanov, and
  Taraskin]{kaznacheev02}
A.~V. Kaznacheev, M.~M. Bogdanov and S.~A. Taraskin, \emph{J.E.T.P.}, 2002,
  \textbf{95}, 57--63\relax
\mciteBstWouldAddEndPuncttrue
\mciteSetBstMidEndSepPunct{\mcitedefaultmidpunct}
{\mcitedefaultendpunct}{\mcitedefaultseppunct}\relax
\EndOfBibitem
\bibitem[Kim \emph{et~al.}(2013)Kim, Shiyanovskii, and Lavrentovich]{kim13}
Y.~K. Kim, S.~V. Shiyanovskii and O.~D. Lavrentovich, \emph{J. Phys.: Condens.
  Matter}, 2013, \textbf{25}, 404202\relax
\mciteBstWouldAddEndPuncttrue
\mciteSetBstMidEndSepPunct{\mcitedefaultmidpunct}
{\mcitedefaultendpunct}{\mcitedefaultseppunct}\relax
\EndOfBibitem
\bibitem[Jamali \emph{et~al.}(2015)Jamali, Behabtu, Senyuk, Lee, Smalyukh,
  van~der Schoot, and Pasquali]{jamali15}
V.~Jamali, N.~Behabtu, B.~Senyuk, J.~A. Lee, I.~I. Smalyukh, P.~van~der Schoot
  and M.~Pasquali, \emph{Phys Rev. E}, 2015, \textbf{91}, 042507\relax
\mciteBstWouldAddEndPuncttrue
\mciteSetBstMidEndSepPunct{\mcitedefaultmidpunct}
{\mcitedefaultendpunct}{\mcitedefaultseppunct}\relax
\EndOfBibitem
\bibitem[Zhou \emph{et~al.}(2017)Zhou, Shiyanovskii, Park, and
  Lavrentovich]{zhou17}
S.~Zhou, S.~V. Shiyanovskii, H.-S. Park and O.~D. Lavrentovich, \emph{Nat.
  Commun.}, 2017, \textbf{8}, 14974\relax
\mciteBstWouldAddEndPuncttrue
\mciteSetBstMidEndSepPunct{\mcitedefaultmidpunct}
{\mcitedefaultendpunct}{\mcitedefaultseppunct}\relax
\EndOfBibitem
\bibitem[DeBenedictis and Atherton(2016)]{debenedictis16}
A.~DeBenedictis and T.~J. Atherton, \emph{Liquid Crystals}, 2016, \textbf{43},
  2352--2362\relax
\mciteBstWouldAddEndPuncttrue
\mciteSetBstMidEndSepPunct{\mcitedefaultmidpunct}
{\mcitedefaultendpunct}{\mcitedefaultseppunct}\relax
\EndOfBibitem
\bibitem[Popa-Nita \emph{et~al.}(1997)Popa-Nita, Sluckin, and
  Wheeler]{popanita97}
V.~Popa-Nita, T.~Sluckin and A.~Wheeler, \emph{J. Phys. II (France)}, 1997,
  \textbf{7}, 1225\relax
\mciteBstWouldAddEndPuncttrue
\mciteSetBstMidEndSepPunct{\mcitedefaultmidpunct}
{\mcitedefaultendpunct}{\mcitedefaultseppunct}\relax
\EndOfBibitem
\bibitem[Frank(1958)]{frank58}
F.~Frank, \emph{Discuss. Faraday Soc.}, 1958, \textbf{25}, 19\relax
\mciteBstWouldAddEndPuncttrue
\mciteSetBstMidEndSepPunct{\mcitedefaultmidpunct}
{\mcitedefaultendpunct}{\mcitedefaultseppunct}\relax
\EndOfBibitem
\bibitem[Selinger(2016)]{selinger16}
J.~V. Selinger, in \emph{Liquid Crystals}, Springer International Publishing,
  2016, pp. 131--182\relax
\mciteBstWouldAddEndPuncttrue
\mciteSetBstMidEndSepPunct{\mcitedefaultmidpunct}
{\mcitedefaultendpunct}{\mcitedefaultseppunct}\relax
\EndOfBibitem
\bibitem[Longa \emph{et~al.}(1987)Longa, Monselesan, and Trebin]{longa87}
L.~Longa, D.~Monselesan and H.~R. Trebin, \emph{Liq. Cryst.}, 1987, \textbf{2},
  769\relax
\mciteBstWouldAddEndPuncttrue
\mciteSetBstMidEndSepPunct{\mcitedefaultmidpunct}
{\mcitedefaultendpunct}{\mcitedefaultseppunct}\relax
\EndOfBibitem
\bibitem[Ball and Majumdar(2010)]{ball10}
J.~M. Ball and A.~Majumdar, \emph{Mol. liq. Cryst.}, 2010, \textbf{525},
  1\relax
\mciteBstWouldAddEndPuncttrue
\mciteSetBstMidEndSepPunct{\mcitedefaultmidpunct}
{\mcitedefaultendpunct}{\mcitedefaultseppunct}\relax
\EndOfBibitem
\bibitem[Schimming \emph{et~al.}(2021)Schimming, Vi\~{n}als, and
  Walker]{schimming21}
C.~D. Schimming, J.~Vi\~{n}als and S.~W. Walker, \emph{J. Comp. Phys.}, 2021,
  \textbf{441}, 110441\relax
\mciteBstWouldAddEndPuncttrue
\mciteSetBstMidEndSepPunct{\mcitedefaultmidpunct}
{\mcitedefaultendpunct}{\mcitedefaultseppunct}\relax
\EndOfBibitem
\bibitem[Schimming and Vi\~{n}als(2020)]{schimming20}
C.~D. Schimming and J.~Vi\~{n}als, \emph{Phys. Rev. E.}, 2020, \textbf{101},
  032702\relax
\mciteBstWouldAddEndPuncttrue
\mciteSetBstMidEndSepPunct{\mcitedefaultmidpunct}
{\mcitedefaultendpunct}{\mcitedefaultseppunct}\relax
\EndOfBibitem
\bibitem[Mottram and Newton(2014)]{mottram14}
N.~J. Mottram and C.~J. Newton, \emph{Introduction to {Q}-tensor theory},
  e-print arXiv:1409.3542v2 [cond-mat.soft], 2014\relax
\mciteBstWouldAddEndPuncttrue
\mciteSetBstMidEndSepPunct{\mcitedefaultmidpunct}
{\mcitedefaultendpunct}{\mcitedefaultseppunct}\relax
\EndOfBibitem
\bibitem[Maier and Saupe(1959)]{maier59}
W.~Maier and A.~Saupe, \emph{I. Z. Naturf.}, 1959, \textbf{14}, 882\relax
\mciteBstWouldAddEndPuncttrue
\mciteSetBstMidEndSepPunct{\mcitedefaultmidpunct}
{\mcitedefaultendpunct}{\mcitedefaultseppunct}\relax
\EndOfBibitem
\bibitem[Walker(2018)]{walker18}
S.~W. Walker, \emph{SIAM J. Sci. Comput.}, 2018, \textbf{40}, C234--C257\relax
\mciteBstWouldAddEndPuncttrue
\mciteSetBstMidEndSepPunct{\mcitedefaultmidpunct}
{\mcitedefaultendpunct}{\mcitedefaultseppunct}\relax
\EndOfBibitem
\bibitem[Brand and Pleiner(1987)]{brand87}
H.~R. Brand and H.~Pleiner, \emph{Phys. Rev. A.}, 1987, \textbf{35}, 3122\relax
\mciteBstWouldAddEndPuncttrue
\mciteSetBstMidEndSepPunct{\mcitedefaultmidpunct}
{\mcitedefaultendpunct}{\mcitedefaultseppunct}\relax
\EndOfBibitem
\bibitem[L\:{o}wen(2010)]{lowen10}
H.~L\:{o}wen, \emph{J. Phys. Consdens. Matter}, 2010, \textbf{22}, 364105\relax
\mciteBstWouldAddEndPuncttrue
\mciteSetBstMidEndSepPunct{\mcitedefaultmidpunct}
{\mcitedefaultendpunct}{\mcitedefaultseppunct}\relax
\EndOfBibitem
\bibitem[Wang \emph{et~al.}(2018)Wang, Upmanyu, and Karma]{wang18}
N.~Wang, M.~Upmanyu and A.~Karma, \emph{Phys. Rev. Materials}, 2018,
  \textbf{2}, 033402\relax
\mciteBstWouldAddEndPuncttrue
\mciteSetBstMidEndSepPunct{\mcitedefaultmidpunct}
{\mcitedefaultendpunct}{\mcitedefaultseppunct}\relax
\EndOfBibitem
\bibitem[Chaikin and Lubensky(1995)]{chaikin95}
P.~M. Chaikin and T.~C. Lubensky, \emph{Principles of Condensed Matter
  Physics}, Cambridge University Press, 1995\relax
\mciteBstWouldAddEndPuncttrue
\mciteSetBstMidEndSepPunct{\mcitedefaultmidpunct}
{\mcitedefaultendpunct}{\mcitedefaultseppunct}\relax
\EndOfBibitem
\bibitem[Lyuksyutov(1978)]{lyuksyutov78}
I.~F. Lyuksyutov, \emph{Sov. Phys. JETP}, 1978, \textbf{48}, 178\relax
\mciteBstWouldAddEndPuncttrue
\mciteSetBstMidEndSepPunct{\mcitedefaultmidpunct}
{\mcitedefaultendpunct}{\mcitedefaultseppunct}\relax
\EndOfBibitem
\bibitem[Koch and Harlen(1999)]{koch99}
D.~L. Koch and O.~G. Harlen, \emph{Macromolecules}, 1999, \textbf{32},
  219--226\relax
\mciteBstWouldAddEndPuncttrue
\mciteSetBstMidEndSepPunct{\mcitedefaultmidpunct}
{\mcitedefaultendpunct}{\mcitedefaultseppunct}\relax
\EndOfBibitem
\bibitem[Wolfsheimer \emph{et~al.}(2006)Wolfsheimer, Tanase, Shundyak, van
  Roij, and Shilling]{wolfsheimer06}
S.~Wolfsheimer, C.~Tanase, K.~Shundyak, R.~van Roij and T.~Shilling, \emph{Phys
  Rev. E.}, 2006, \textbf{73}, 061703\relax
\mciteBstWouldAddEndPuncttrue
\mciteSetBstMidEndSepPunct{\mcitedefaultmidpunct}
{\mcitedefaultendpunct}{\mcitedefaultseppunct}\relax
\EndOfBibitem
\bibitem[Vitral \emph{et~al.}(2021)Vitral, Leo, and Vi{\~n}als]{re:vitral21}
E.~Vitral, P.~H. Leo and J.~Vi{\~n}als, \emph{Soft Matter}, 2021, \textbf{17},
  6140--6159\relax
\mciteBstWouldAddEndPuncttrue
\mciteSetBstMidEndSepPunct{\mcitedefaultmidpunct}
{\mcitedefaultendpunct}{\mcitedefaultseppunct}\relax
\EndOfBibitem
\end{mcitethebibliography}
\bibliographystyle{rsc} 

\end{document}